\documentclass{pasj01}
\usepackage{bm}
\draft 
\Received{$\langle$reception date$\rangle$}
\Accepted{$\langle$acception date$\rangle$}
\Published{$\langle$publication date$\rangle$}

\begin{document}

\title{Measurement of Vector Magnetic Field in a Flare kernel with a Spectropolarimetric Observation in He \emissiontype{I} 10830 \AA}
\author{Tetsu Anan\altaffilmark{1,2,*}, Takurou Yoneya\altaffilmark{3,4}, Kiyoshi Ichimoto\altaffilmark{2,5},
Satoru Ueno\altaffilmark{2}, Daikou Shiota\altaffilmark{6}, Satoshi Nozawa\altaffilmark{3},
Shinsuke Takasao\altaffilmark{7}, Tomoko Kawate\altaffilmark{8}} %
\altaffiltext{1}{National Solar Observatory, 22 Ohi'a Ku, Makawao, HI, 96768, USA}
\altaffiltext{2}{Astronomical Observatory, Kyoto University, Kamitakara, Gifu, 506-1314, Japan}
\altaffiltext{3}{Department of Science, Ibaraki University, 
2-1-1 Bunkyo, Mito, Ibaraki 310-8512, Japan}
\altaffiltext{4}{Techmatrix.co, 3-11-23, Mita, Minato-ku, Tokyo, 108-0073, Japan }
\altaffiltext{5}{National Astronomical Observatory of Japan, 2-21-1 Osawa, Mitaka, Tokyo, 181-8588, Japan}
\altaffiltext{6}{Applied Electromagnetic Research Institute, National Institute of Information and Communications Technology, 4-2-1 Nukui-Kitamachi, Koganei, Tokyo 184-8795, Japan}
\altaffiltext{7}{Department of Physics, Nagoya University, Nagoya, Aichi, 464-8602, Japan}
\altaffiltext{8}{Institute of Space and Astronautical Science, Japan Aerospace Exploration Agency, Sagamihara, Kanagawa 252-5210, Japan}

\email{tanan@nso.edu}


\maketitle

\begin{abstract}
A flare kernel associated with a C4 class flare was observed in a spectral window including the He \emissiontype{I} triplet 10830 \AA\, and Si \emissiontype{I} 10827\AA\, with a spectropolarimeter on the Domeless Solar Telescope at Hida Observatory on August 9th, 2015.
Observed Stokes profiles of the He \emissiontype{I} triplet in the flare kernel in its post-maximum phase are well reproduced through inversions considering the Zeeman and the Paschen-Back effects with a three slab model of the flare kernel, in which two slabs having up and downward velocities produce emissions and one slab produces an absorption.
The magnetic field strength inferred from the emission components of the He \emissiontype{I} line is 1400 G, which is significantly stronger than 690 G that is observed at the same location in the same line 6.5 hours before the flare.
In addition, photospheric magnetic field vector derived from the Si \emissiontype{I}10827\AA\,is similar to that of the flare kernel.
To explain this results, we suggest that the emission in the He \emissiontype{I} triplet during the flare is produced in the deep layer, around which bombardment of non-thermal electrons leads to the formation of a coronal temperature plasma.
Assuming a hydrogen column density at the location where the He \emissiontype{I} emissions are formed, and a power-law index of non-thermal electron energy distribution, we derived the low-energy cutoff of the non-thermal electron as 20 - 30 keV, which is consistent with that inferred from hard X-ray data obtained by RHESSI.
\end{abstract}

\section{Introduction}
\label{sec.intro}

Solar flares are sudden brightenings in the entire electromagnetic solar spectrum, from radio to $\gamma$-rays \citep{fletcher11}.
Their phenomenological aspects are explained as the release of magnetic energy stored in the solar corona through magnetic reconnections (\cite{carmichael64}, \cite{sturrock66}, \cite{hirayama74}, \cite{kopp76}, \cite{nishio97}, \cite{melrose97}, \cite{shibata99}).
Evaluation of amount of the magnetic energy that is converted to other forms of energy, such as non-thermal electrons, heat, and plasma motions, is an important subject for understanding the mechanism of the energy release \citep{emslie12, aschwanden17}.  

Solar chromosphere dynamically responds to solar flares through the energy injection from the corona by thermal conduction and/or non-thermal particles \citep{nagai80, somov81, somov82, ichimoto84}.
For the thermal conduction, the transition region between the corona and the chromosphere will propagate downward due to the heating and subsequent evaporation of the upper part of the chromosphere.
For the non-thermal particles, temperature may increase at a location in the chromosphere at which the non-thermal electrons penetrate and dissipate large part of their kinetic energy \citep{nagai84}.
Therefore, the dynamical responses of the chromosphere to the flare will be different against the different heating mechanisms, and the investigation of it will provide us crucial information about the mechanism of energy transport in flares and their energetics.

The polarization of photospheric or chromospheric spectral lines is produced by magnetic fields through the Zeeman and Hanle effects.
Furthermore, when energetic particles with anisotropic velocity distribution, e.g., the accelerated charged particles and heat flux, penetrate into the chromosphere, collisions with ambient atoms will produce the impact polarization depending on the kinetic energy and the velocity distribution of the particles \citep{henoux83a, henoux83b}.
Radiation anisotropy due to plasma inhomogeneities at the boundary of the flare ribbons also can induce scattering polarization in the spectral lines \citep{stepan13}.
A number of works have reported observations of the polarization in spectral lines associated with flares, and they have discussed the origin of the polarization \citep{henoux83a, hanaoka03, xu05, judge15}.

According to the current flare models (see e.g. \cite{shibata11}), flare ribbons observed in the chromosphere correspond to foot points of the reconnecting and restructuring coronal magnetic loops.
Hence, measurements of the magnetic field strength in the flare ribbons allow us to estimate the reconnecting magnetic flux, and the released magnetic energy \citep{isobe02, asai04}.
In addition, inferred magnetic field variations in photosphere during flares \citep{wang92, sudol05} are interpreted as a result of coronal field restructuring \citep{anwar93, hudson08}.

Through spectro-polarimetric observations of chromospheric line, \citet{penn95} and \citet{harvey12} derived magnetic field component along the line of sight from circular polarization, i.e. Stokes ${\it V}$, of the He \emissiontype{I} 10830 \AA\,and Ca \emissiontype{II} 8542 \AA\,, respectively.
\citet{kleint17} analyzed Stokes ${\it V}$ imaging spectro-polarimetric data in the Ca \emissiontype{II} 8542 \AA\, and found that a significant decrease and increase of chromospheric magnetic field component along the line of sight during a flare, and its variation is twice as large as that in the underlying photosphere.
They attributed the change of the chromospheric magnetic field to the coronal field restructuring during the flare.
However, since the direction of the magnetic fields was not measured, it remains unresolved if the coronal field restructuring was the major cause of the large changes in the observed fields.
\citet{kuridze18} applied spectro-polarimetric inversions \citep{socas-navarro15} to Ca \emissiontype{II} 8542 \AA\,full-Stokes profiles of a flare ribbon, and suggested that the line in the flaring atmosphere is formed in the deeper layers than in the quiet Sun.
They also present strong $\sim 1000$ - $1500$ G transverse component of the magnetic field in a portion of the flare ribbon where linear polarization is the strongest. 


This paper presents the first measurement of magnetic field vectors, i.e., directions and strengths, in a flare kernel with a spectro-polarimetric observation of the He \emissiontype{I} 10830 \AA.
In the following sections, we describe the details of the observation (section \ref{sec.observation}), the results on the magnetic field vectors derived from Stokes profiles of the He \emissiontype{I} 10830 \AA\,in the flare kernel (section \ref{sec.result}), and finally we discuss possible interpretations and summarize our findings (section \ref{sec.sammarry}).

\section{Observation}
\label{sec.observation}

\begin{figure}
 \begin{center}
  \includegraphics[width=12cm]{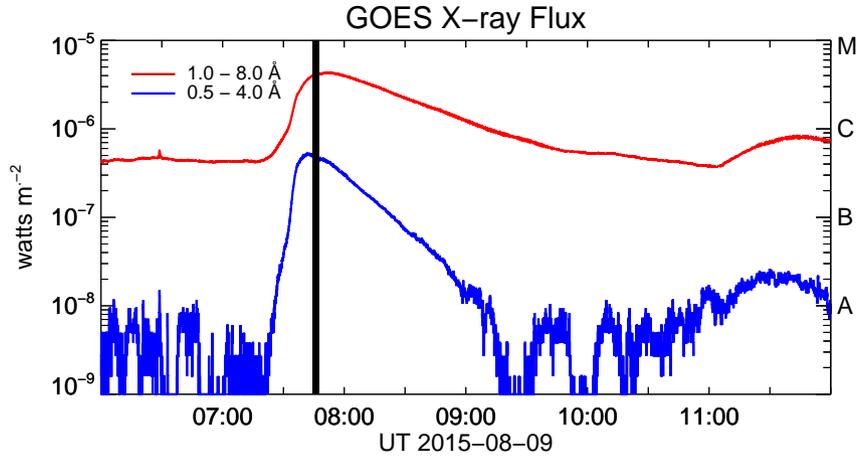}
 \end{center}
 \caption{
 X-ray light curves measured by GOES at 6:00-12:00 UT on August 9, 2015.
 The red and blue solid lines indicate the intensity between 1.0 - 8.0 \AA, and 0.5 - 4.0 \AA, respectively.
 The black vertical line marks the time when flare kernel was observed with the spectropolarimeter.}
 \label{fig.goes}
\end{figure}
\begin{figure}
 \begin{center}
  \includegraphics[width=12cm]{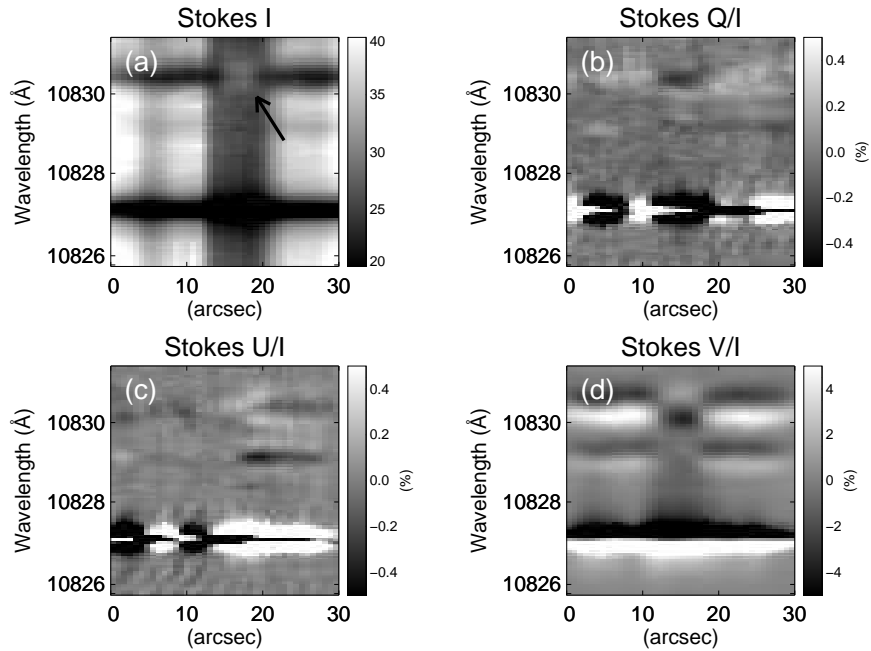}
 \end{center}
 \caption{
Full Stokes spectra of the flare kernel on the sunspot in He \emissiontype{I} triplet and Si \emissiontype{I} 10827 \AA.
A black arrow points the emission of the flare kernel in the He \emissiontype{I} triplet. 
}
 \label{fig.stokes_spectra}
\end{figure}

A C4 class flare occurred on August 9th, 2015 in the active region NOAA 12396 at the heliocentric coordinates of S\timeform{19D}W\timeform{14D}.
Figure \ref{fig.goes} shows the X-ray light curves of the flare measured by Geostationary Operational Environmental Satellite (GOES).
We observed the active region in He \emissiontype{I} 10830 \AA\, and Si \emissiontype{I} 10827 \AA\, with the spectropolarimeter equipped on the Domeless Solar Telescope \citep{anan12, nakai85} at Hida observatory.
Two dimensional maps of the active region were obtained by scanning the slit of the spectropolarimeter with a scan step of 2 arcsec during and 6.5 hours before the flare.
At each slit position, 100 frames of polarization-modulated spectral images in both orthogonal polarizations send from the polarization analyzer were recorded in 20 sec, and it took 60 min for obtaining the field of view of 128 $\times$ 240 arcsec$^2$.
The slit width is 1.28 arcsec and the spatial samplings along the slit is 0.6 arcsec pixel$^{-1}$.
The spectral sampling is 37 m\AA\, pixel$^{-1}$.

\begin{figure}
 \begin{center}
  \includegraphics[width=12cm]{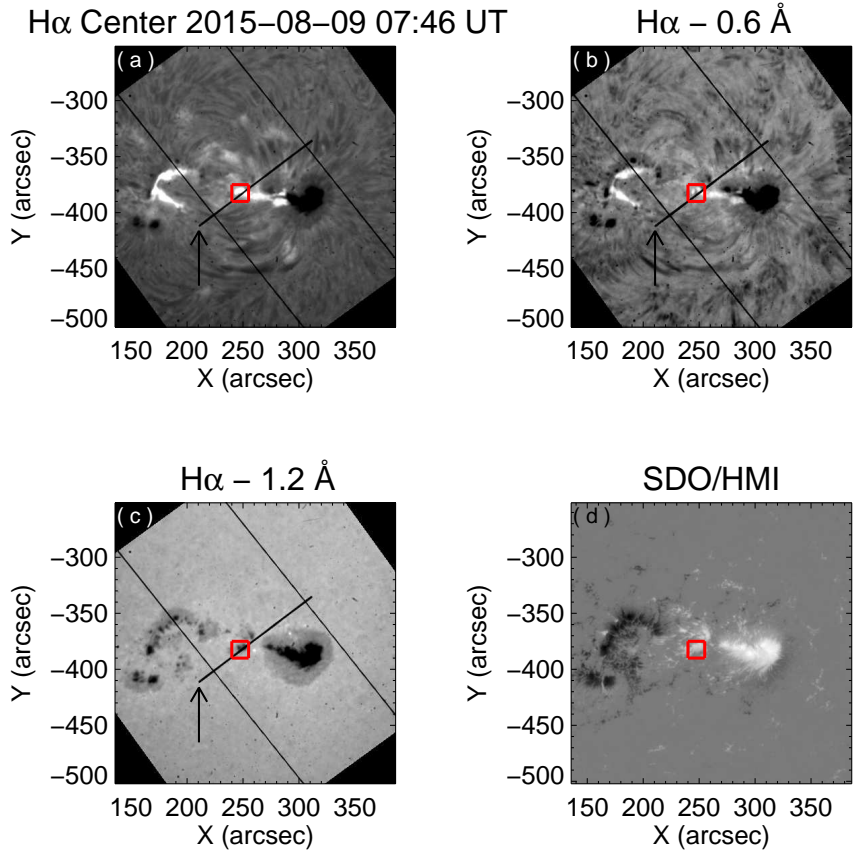}
 \end{center}
 \caption{
 Slit-jaw images in (a) H$\alpha$ center, (b) H$\alpha -$ 0.6 \AA, (c) H$\alpha -$ 1.2 \AA, and (d) a photospheric magnetogram at the same time when the emissions of the flare kernels were measured.
 The black lines pointed by arrows are the slit of the spectropolarimeter.
  The two parallel lines are hair lines.
  The red boxes mark the regions where the flare kernel was observed.
  }
  \label{fig.his}
\end{figure}
\begin{figure}
 \begin{center}
  \includegraphics[width=10cm]{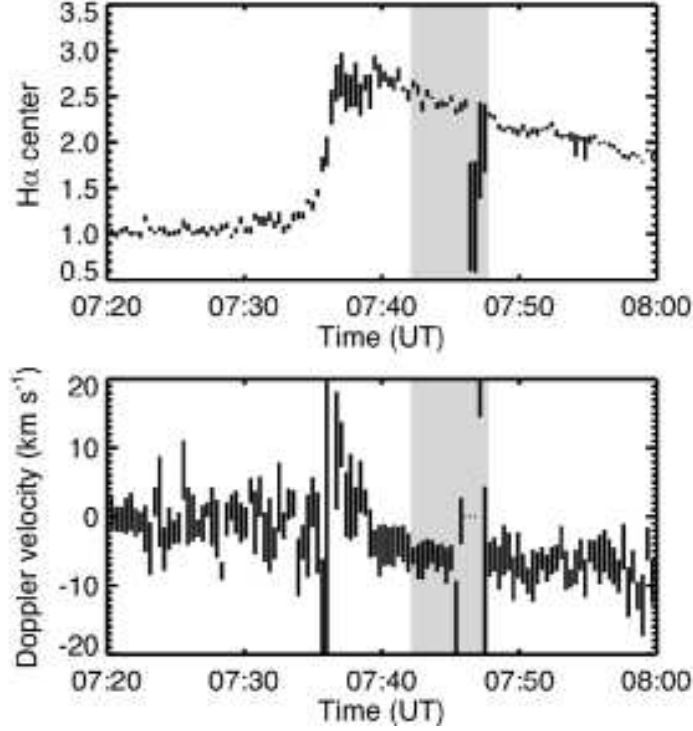}
 \end{center}
 \caption{
 Light curve in the H$\alpha$ center and temporal variation of the Doppler velocity of the H$\alpha$ emitted from the sunspot.
 The gray region indicates the time when emissions of flare kernels in the He \emissiontype{I} 10830 \AA\, were observed.
}
 \label{fig.his_light_curve}
\end{figure}

Emissions of the He \emissiontype{I} 10830 \AA\, were observed on a sunspot in the active region from 7:43 UT to 7:47 UT.
Figure \ref{fig.stokes_spectra} shows full Stokes spectra that we obtained when the slit of the spectrograph was placed on the flare kernel.
The full Stokes spectra were calibrated for instrumental polarization using a predetermined Mueller matrix of the telescope \citep{anan12}.
Seeing induced cross-talks can be reduced by using the both orthogonal polarizations split by the polarization analyzer in the reduction process.
However, the Stokes spectra still show signatures of remaining cross-talks among the Stokes parameters.
The cross-talks from Stokes ${\it I}$ to ${\it Q}$, ${\it U}$, and ${\it V}$ were corrected by subtracting a bias in the continuum in ${\it Q}$, ${\it U}$, and ${\it V}$, so that there is no spurious polarization in the continuum, in which polarization degree should be less than $10^{-6}$ on the solar disk \citep{stenflo05}.
After removing fringes by filtering them out in Fourier space, we applied a technique proposed by \citet{kuhn94} to correct the cross-talks between the linearly and circularly polarized states.
The technique is based on assumptions that the Stokes-${\it Q}$ and ${\it U}$ profiles in the Si \emissiontype{I} 10827 \AA\,are symmetric and its Stokes-${\it V}$ profiles are anti-symmetric.
The polarization sensitivity (rms noise level) in the Stokes ${\it Q}$, ${\it U}$, and ${\it V}$ are, respectively, estimated to be $3 \times 10^{-4}$, $2 \times 10^{-4}$, and $7 \times 10^{-4}$ as the standard deviation of polarization degrees in a continuum part of the spectra.

Slit-jaw images at several wavelengths in the H$\alpha$ line are obtained during the spectro-polarimetric scan through a tunable Lyot filter with a passband width of 0.5\AA\,(FWHM) and a 1k $\times$ 1k CCD camera.
Figure \ref{fig.his} shows the slit-jaw images in H$\alpha$ center, H$\alpha$ $-0.6$ \AA, and  H$\alpha$ $-1.2$ \AA\,taken at the same time when the emissions of the flare kernels were measured in the He \emissiontype{I} 10830 \AA.
The flare kernel is located in a sunspot and a part of a flare ribbon, which appear in the slit-jaw images of H$\alpha$ center and H$\alpha$ $-0.6$ \AA.
The photospheric magnetogram were taken by the Helioseismic and Magnetic Imager (HMI, \cite{scherrer12}) onboard the Solar Dynamics Observatory (SDO).

\begin{figure}
 \begin{center}
  \includegraphics[width=14cm]{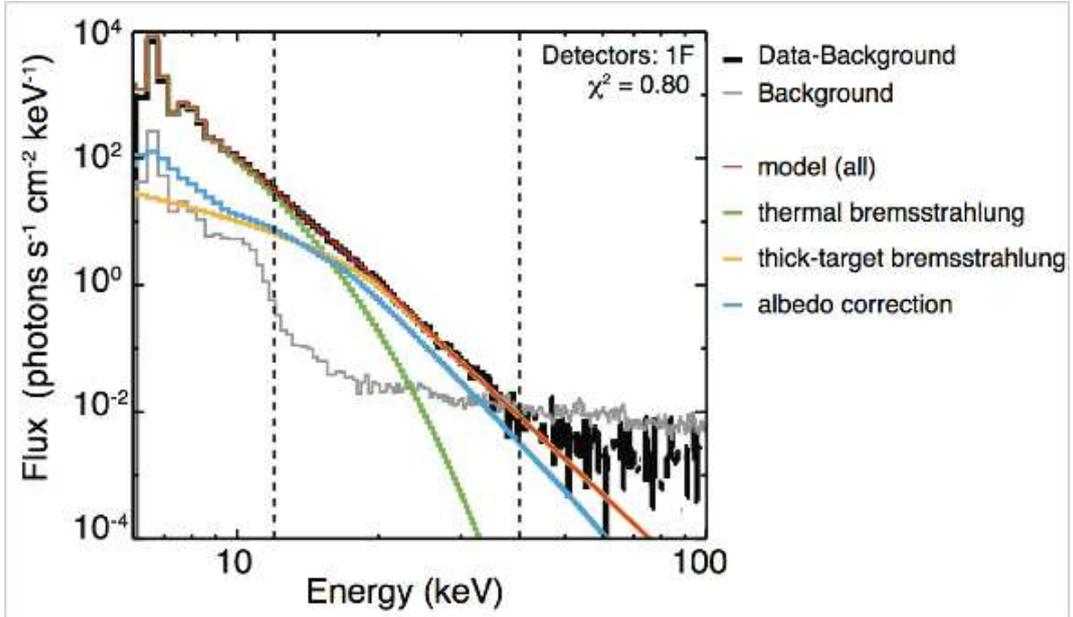}
 \end{center}
 \caption{
  A hard X-ray spectrum in the impulsive phase of the flare observed by RHESSI detecter 1.
  Background, background-subtracted data, and the fit are plotted by gray, black, and red curves, respectively.
  Best-fit models of thermal bremsstrahlung (green), non-thermal thick-target bremsstrahlung (orange), and back scattering (albedo correction, blue) are also shown. Vertical dashed lines indicate the boundaries of the fitting range between 12 and 40 keV. 
 The inferred electron power-law index and upper limit of the low-energy cutoff of the non-thermal electron energy distribution are $7.8 \pm 0.7$ and $21.7 \pm 1.4$ keV, respectively.
}
 \label{fig.hxr_spectrum}
\end{figure}

Figure \ref{fig.his_light_curve} shows a light curve of H$\alpha$ center and the Doppler velocity of the H$\alpha$ at the flare kernel. The period when we observed the emission in HeI 10830 \AA\, is shown by a vertical shade.
After the dark-frame correction, the intensities of the slit-jaw images are normalized to spatially averaged intensity of quiet regions in the field of view for each wavelength.
We fit the H$\alpha$ emission profiles sampled at the H$\alpha$ center, H$\alpha$ $\pm0.3$ \AA, H$\alpha$ $\pm0.6$ \AA, H$\alpha$ $\pm0.9$ \AA, and  H$\alpha$ $\pm1.2$ \AA\, with a Gaussian function and the Doppler velocity is determined from its peak position. 
The figure indicates that the red shift with a speed of 10 km s$^{-1}$ was detected in the rising phase of the H$\alpha$ emission, and the spectro-polarimetric measurement of the flare kernel in the He \emissiontype{I} 10830 \AA\, was obtained after the peak of the H$\alpha$ intensity.
The slit scan over the whole active region took one hour.
Because of this long scanning period, we cannot study the time evolution of the Stokes profiles in the flare kernel in the flare onset.

Reuven Ramaty High Energy Solar Spectroscopic Imager  (RHESSI, \cite{lin02}) also observed the flare in hard X-ray range until 7:46 UT.
Figure \ref{fig.hxr_spectrum} shows the spectra of the photon flux integrated over the period of 07:33:44-07:37:40 UT, in which the photon counts integrated over 25-50 keV were above the background level.
We performed spectral fitting using the ${\it OSPEX}$ procedure in the Solar SoftWare (SSW) with an electron source model, in which hard X-ray spectrum consists of components created by thermal bremsstrahlung and non-thermal thick-target bremsstrahlung by electrons with a single power-law energy distribution while effects of the albedo and the pileup modulation are corrected.
According to the fitting results of data by RHESSI detectors 1, 5, 7, and 8, the inferred power-law index and upper limit of the low-energy cutoff of the non-thermal electron energy distribution are $7.8 \pm 0.7$ and $21.7 \pm 1.4$ keV, respectively.

%

\section{Result}
\label{sec.result}

\begin{figure}
 \begin{center}
  \includegraphics[width=14cm]{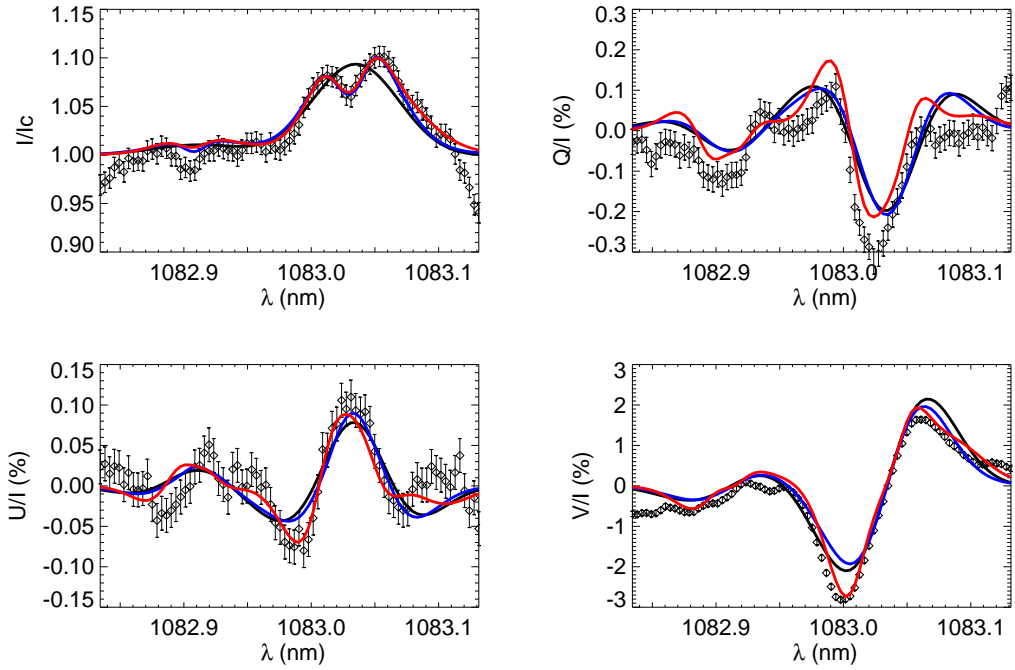}
 \end{center}
 \caption{
Observed Stokes profiles of the flare kernel in the He \emissiontype{I} 10830 \AA\, (black diamonds), and inversion results for single (black solid line), double (blue solid line), and triple slabs models (red solid line).
}
 \label{fig.fitting}
\end{figure}

Full Stokes profiles of the He \emissiontype{I} 10830 \AA\,at the flare kernel are shown in Figure \ref{fig.fitting}.
Stokes ${\it Q}$, ${\it U}$, and ${\it V}$ present a hint of profiles produced by the Zeeman effect, while the Stokes ${\it I}$ shows two peaks of the emission.
Note that the sign of the Stokes ${\it V}$ profile of the kernel in the He \emissiontype{I} 10830 \AA\, is opposite to those of surrounding regions and in the Si  \emissiontype{I} 10827 \AA\, (figure \ref{fig.stokes_spectra}).
In weak field approximation, Stokes ${\it V}$ can be written as 
\begin{equation}
{\it V} \sim - g \lambda^2 B_{{\rm LOS}} \frac{\partial {\it I}}{\partial \lambda},
\end{equation}
where $g$ is the Land${\rm \acute{e}}$ factor, $\lambda$ is the wavelength, and $B_{{\rm LOS}}$ is the line of sight component of the magnetic field (see chapter 11.9 of \cite{stenflo94}).
The opposite sign of the Stokes ${\it V}$ profile is 
due to that the Stokes-${\it V}$ signal is produced by the Zeeman effect of the emission component, in which the $\partial {\it I} / \partial \lambda$ turns to the opposite sign of that of the absorption profiles.
Our inversion code can properly handle the Zeeman effect of the emission component.

\begin{longtable} {* {9} {l}}
\caption{Inverted physical parameters}\label{tbl.result1}
\hline
      time (UT) & line                          & B (G) & $\gamma$ ($^{\circ}$) & $\phi$ ($^{\circ}$)           & v (km s$^{-1}$) & $\Delta \lambda$ (\AA) & S    & $\tau$    \\ 
\hline
\endfirsthead
     \multicolumn{9}{l}{\footnotemark[$*$] Physical parameters of each slab in the triple slab model. The first slab produces an absorption ($S<1$),} \\
     \multicolumn{9}{l}{while the second and the third slabs produce emissions ($S>1$) with an upward and downward velocity.}\\
     \multicolumn{9}{l}{\footnotemark[$\dagger$] Optical depths are fixed in the fitting} \\
  \endfoot
\hline
      01:13           & He \emissiontype{I} & 680$\pm$40       & 15$\pm$9   & 125$\pm$75& 1.6$\pm$0.1                                       & 0.304$\pm$0.005        &  0.21$\pm$0.01    &   0.4\footnotemark[$\dagger$]   \\      
      01:13           & Si \emissiontype{I}  & 1070$\pm$20     & 46$\pm$2     & 268$\pm$2&  1.2$\pm$0.1                                      & 0.053$\pm$0.005         &  \multicolumn{2}{c}{0.4 + 0.6$\tau$}                          \\ \hline    
      \footnotemark[$*$]07:45 (flare, L1)& He \emissiontype{I}  & 360 $\pm$ 100  &  60$\pm$40 & 190$\pm$115& -0.2$\pm$0.2     & 0.14$\pm$0.02   & $<$0.68   &     0.4\footnotemark[$\dagger$]             \\     
      \footnotemark[$*$]07:45 (flare, L2)& He \emissiontype{I} & 1380$\pm$ 120  &  51$\pm$5   & 277$\pm$5    &  -1.2$\pm$0.3    & 0.26$\pm$0.02             & $>$1.42   &    0.4\footnotemark[$\dagger$]            \\       
      \footnotemark[$*$]07:45 (flare, L3)& He \emissiontype{I} & 1380$\pm$ 100  &  50$\pm$6   & 271$\pm$8    &     7$\pm$4        & 0.38$\pm$0.09             & $>$1.23   &    0.4\footnotemark[$\dagger$]          \\   
      07:45 (flare)& Si \emissiontype{I}  & 1470$\pm$ 20   &    57$\pm$1  & 268$\pm$1 & 1.3$\pm$0.1                                       & 0.051$\pm$0.005         &     \multicolumn{2}{c}{0.07 + 0.93$\tau$}                           \\         \hline            
\end{longtable}
%
%
%
%
%
\begin{figure}
 \begin{center}
  \includegraphics[width=14cm]{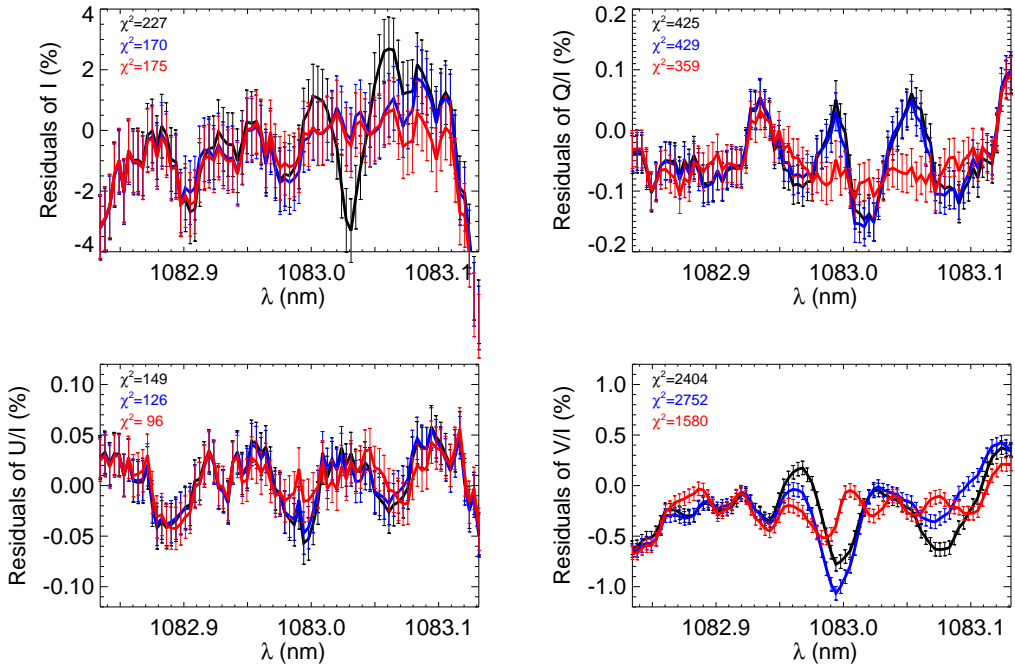}
 \end{center}
 \caption{
Residuals and the chi-square values, $\chi^2$, of the fittings for single (black), double (blue), and triple slabs models (red).
}
 \label{fig.fitting_rms}
\end{figure}


We fit the observed Stokes spectra with theoretical profiles computed with a radiative transfer equation \citep{iniesta03} taking into account the Zeeman and Paschen-Back effects in the presence of the magnetic field by applying Levenberg-Marquardt least squares minimization. 
The atmospheric models used here are based on a constant-property slab model for the He \emissiontype{I} 10830 \AA\,and the Milne-Eddington model for the Si \emissiontype{I} 10827 \AA , respectively.
The constant-property slab is a simple and suitable model for calculating Stokes profiles of the He \emissiontype{I} 10830 \AA\, \citep{ramos08}.
Optical depth, $\tau$, is fixed to 0.4 in the fitting for the He \emissiontype{I} 10830 \AA\,of the flare kernel, because it cannot be well determined simultaneously with the source function, since in optically thin case, the intensity of the emission is only proportional to the product of $\tau$ and the source function.

Considering that the observed Stokes {\it I} profile of the flare kernel has two peaks, we also perform fittings using multi-slab model, in which the observed Stokes profiles are sum of profiles produced by each slab.
In case of triple slab model, a Stokes parameter, $p$, is calculated as $p=\sum_{i=1}^3 (f_{i} p_{i})$ under the constraint of $\sum_{i=1}^3 f_{i} = 1$, where $p_{i}$ and $f_{i}$ are a Stokes parameter and a filling factor of the $i$th slab, respectively.

Figure \ref{fig.fitting} shows the results of inversion of He \emissiontype{I} 10830 \AA\,with a single, double, and triple slab models overlaid on the observed Stokes profiles of the He \emissiontype{I} 10830 \AA\, in the flare kernel.
The residuals of the fittings in three cases are presented in figure \ref{fig.fitting_rms} with the chi-square values.
There remain some systematic patterns in the residuals with a scale of line width. They may be attributed to remaining fringes or presence of gradients of physical quantities in the flare kernel.
However overall characteristics of the Stokes profiles are best represented by the triple slab models with the least chi-square values for the Stokes ${\it Q}$, ${\it U}$ and ${\it V}$.

The two peaks in the Stokes-${\it I}$ profile cannot be reproduced by a single component model. 
A fitting with the double slab model yields that the Stokes-${\it I}$ profile is almost consistent with a wide emission and a narrow absorption at the center of 1083.3 nm.
However, the double slab model is not able to reproduce the asymmetric Stokes profiles, for example, the difference between the peaks of positive annd negative lobes of the Stokes-${\it V}$ profile cannot be reproduced, as presented by the blue curve in figure \ref{fig.fitting}.
In order to explain the observed Stokes profiles, we need at least two emission components with different Doppler velocities and one absorption component.
It is found in figure \ref{fig.fitting} (red curve) that the triple slab model well reproduces not only the two peaks of the Stokes-${\it I}$ profile, but also the asymmetry of the Stokes-${\it Q}$, -${\it U}$, and -${\it V}$ profiles.

%


Tables \ref{tbl.result1} summarize the results of the fitting with the triple slab model for He \emissiontype{I} 10830 \AA\,for the flare kernel together with the results for Si \emissiontype{I} 10827 \AA\,and for the same location but 6.5 hours before the flare.
In the triple slab model for the flare kernel, in which each slab is assumed to be optically thin, the contribution from a slab on the emergent Stokes profiles is proportional to $f_{i}(S_{i}-1)$, where $S_{i}$ is the source function of $i$th slab normalized to that of the background.
Therefore, it is not possible to determine $f_{i}$ and $S_{i}$ separately, but only determined from the fitting is the product, $f_{i}(S_{i}-1)$.
The results of the best fitting shown in figure \ref{fig.fitting} are $f_{1}(S_{1}-1)=-0.32$, $f_{2}(S_{2}-1)=0.42$, and $f_{3}(S_{3}-1)=0.23$.
From $f_{i}>0$, it is obvious that $S_{1} < 1$ (source of an absorption component of the Stokes profile) and $S_{2}$, $S_{3}>1$ (source of emission components of the Stokes profile).
More precisely, we can constrain each source function as follows; $S_{1} < 0.65$ from $f_{1} > 0$, $S_{2} > 1.42$ from $f_{2} < 1$, and $S_{3} > 1.23$ from $f_{3} < 1$.
It is noted that the first component producing the absorption is almost stationary ($V \sim 0$), while the second and third components producing the emission have upward and downward velocities, respectively.

Other parameters obtained from the fitting and shown in Tables \ref{tbl.result1} are the magnetic field strengths, $B$, inclinations of magnetic field vectors with respect to the local solar vertical, $\gamma$, azimuth angles about the local solar vertical with respect to the east direction on the sun, $\phi$, and line width, $\Delta \lambda$.
In non-flaring regions the spectral lines of He \emissiontype{I} and Si \emissiontype{I} are formed in the upper chromosphere and the upper photosphere, respectively \citep{avrett94, orozco17}.
Therefore the result of the fitting at the same location 6.5 hours before the flare infers that the magnetic field strength decrease with height above the umbra.
It is interesting to find that during the flare the magnetic field strengths derived from the He \emissiontype{I} emissions are significantly (four times) larger than that of the He \emissiontype{I} absorption, and they are two times larger than that inferred from the He \emissiontype{I} Stokes profiles before the flare occurrence.
In addition, the field orientations of the He \emissiontype{I} emissions are similar to that derived from the Si \emissiontype{I} in the flare kernel.

\section{Discussion \& Summary}
\label{sec.sammarry}

We present the first measurement of the magnetic field vectors on the flare kernel with a spectro-polarimetric observation of the He \emissiontype{I} 10830 \AA.
The observed Stokes profiles of the flare kernel in the He \emissiontype{I} 10830 \AA\,are well reproduced by the triple slab model that consists of two emission components with an upward and downward Doppler velocities and one absorption component.
Another important findings are that the magnetic field strengths of the emission components are two times larger than that inferred from the Stokes profiles in the He \emissiontype{I} 10830 \AA\, before the flare occurrence, and four times larger than that of the absorption component.

\begin{figure}
\begin{center}
 \includegraphics[width=14cm]{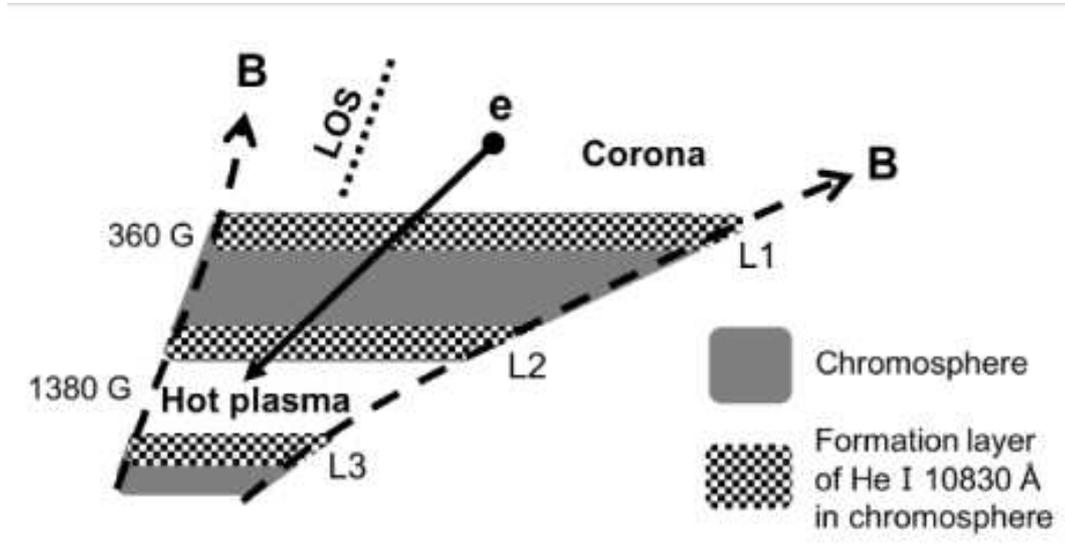}
 \end{center}
 \caption{
  An interpretation of the inferred magnetic field vectors on the flare kernel.
  The dotted lines indicate the line of sight.
  }
  \label{fig.interpret}
\end{figure}

One of possible interpretations for this result is that the emissions of the He \emissiontype{I} 10830 \AA\,originate from the heating in the deep layer of the flare kernel due to bombardment of non-thermal electrons (figure \ref{fig.interpret}).
When non-thermal electrons penetrates the chromosphere, they will stop at a depth determined by their kinetic energy and the ambient density, and it will lead to the formation of a coronal temperature plasma in the energy-deposited layer \citep{nagai84, tei18}.
As a result, the He \emissiontype{I} 10830 \AA\,will be produced either ways by direct excitation by the non-thermal electrons in the chromosphere, 
collisional excitation by thermal electrons in compressed chromosphere next to the coronal temperature plasma, or the photoionization by EUV radiation with a wavelength $< 504$ \AA\, emitted from the coronal temperature plasma \citep{ding05, zeng14}.
The source of the observed two emission components with an upward and downward velocity (the fourth and fifth rows of the table \ref{tbl.result1}) can be attributed to the top and bottom boundary layers of the coronal temperature plasma (L2 and L3 in the Figure \ref{fig.interpret}), and the source of the absorption component (the third row of the table \ref{tbl.result1}) can be in the upper chromosphere, through which the non-thermal electron beam propagates without dissipation (L1 in the figure \ref{fig.interpret}).  

The magnetic field strength may decrease with the height in the flare kernel, because the field strength in the upper photosphere (the sixth row of the table \ref{tbl.result1}) is four times larger than that in the upper chromosphere
(the third row of the table \ref{tbl.result1}).
The magnetic field strengths of the observed two emission components of the He \emissiontype{I} 10830 \AA\, are also much larger than that in the upper chromosphere during and 6.5 hours before the flare (compare the first, fourth, fifth, and sixth rows of the table \ref{tbl.result1}).
It is consistent with the scenario in which the emissions are produced in the deep layer.

The depth where the non-thermal electrons deposit the most energy depends on the low-energy cutoff in their energy spectrum \citep{emslie78}.
We can estimate the low-energy cutoff by assuming the hydrogen column density in the pre-flare chromosphere \citep{allred15}.
In non-flaring regions, the He \emissiontype{I} 10830 \AA\, line is supposed to be formed in the upper chromosphere \citep{avrett94}, and the difference of the formation heights of the He \emissiontype{I} 10830 \AA\,and the Si \emissiontype{I} 10827 \AA\,lines are estimated to be approximately 1000 km in sunspot umbrae from the analysis of propagating waves \citep{centeno09}.
If we assume that the non-thermal electron beam with the power-law index of 7.8 (section \ref{sec.observation}) penetrate into the chromosphere and stop at the depth with a hydrogen column density of $1$ - $3 \times 10^{20}\,{\rm cm^{-2}}$, which corresponds to the depth of 1000 km below the transition region according to the umbral model of \citet{maltby86}, the low-energy cutoff can be derived as $\sim$20-30 keV (see figures 5, 6, and 9 of  \cite{allred15}).
Because the Si \emissiontype{I} 10827 \AA\,did not show emission, the hydrogen column density of $1$ - $3 \times 10^{20}\,{\rm cm^{-2}}$, or the derived lower cutoff energy, will be the maximum possible value.  
The hard X-ray spectral fit yields that the upper limit of the low-energy cutoff is $21.7 \pm 1.4$ keV (section \ref{sec.observation}), and it is consistent with the result here.

It must be noted that there may be other possible scenarios, such as the coronal field restructuring and change of line formation height due to thermal conduction. 
In these scenarios, three components in the triple slab model can be interpreted as three different unresolved magnetic loops.

The observed Stokes-${\it Q}$ and ${\it U}$ profiles are qualitatively different from those reported in a X-class flare by \citet{judge15}.
They concluded that the linear polarization in He \emissiontype{I} 10830 \AA\,in the X-class flare are naturally explained with the scattering polarization under an anisotropic radiation field, while we successfully reproduced our observed profiles by the Zeeman effect without taking the scattering polarization into account.
It might be possible that the kernels observed in these two flares are located in different conditions of the radiation fields and the magnetic fields at the He \emissiontype{I} 10830 \AA\, formation layer.

Finally we like to stress that detailed measurements of temporal and spacial distributions of magnetic fields and plasma dynamics in flare kernels have a potential to provide valuable diagnoses of the non-thermal electrons in flares, and such observations at much higher spatial resolutions are expected to be performed by new generation instruments, such as Daniel K. Inouye Solar Telescope (DKIST, \cite{rimmele05}).

\bigskip
\begin{ack}
This work was supported by JSPS KAKENHI Grant Number 22244013, JP16H01177, JP15K17609, 26560079, and 16H03955.
We thank staffs and students of Ibaraki University and Kyoto University, especially Dr. A. Asai, Dr. S. Nagata, Dr. T. T. Ishii, Ms. A. Tei, and Mr. T. Nakamura.
\end{ack}

{}

\end{document}